\documentclass[conference]{IEEEtran}
\IEEEoverridecommandlockouts

\usepackage{amsmath,amssymb,amsfonts}
\usepackage{algorithmic}
\usepackage{graphicx}
\usepackage{textcomp}

\DeclareGraphicsExtensions{.pdf,.jpg,.png,.jpeg}

\usepackage[
style=ieee,doi=false,isbn=false,url=false,eprint=false,maxnames=4
]{biblatex}
\title{{\normalsize Author preprint July 2026 \\ \vspace{-1em}
To appear in the Probabilistic Methods Applied to Power Systems (PMAPS) 2026, Salt Lake City, Utah}\\
Measuring outage resilience in a distribution system with the number of outages in large events
\thanks{
Support from USA NSF grants 2153163, 2429602, Argonne National Laboratory, and PSerc project S110 is gratefully acknowledged.}}

\author{\IEEEauthorblockN{Arslan Ahmad}
\IEEEauthorblockA{Iowa State University\\
arslan@iastate.edu}
\and
\IEEEauthorblockN{Ian Dobson}
\IEEEauthorblockA{Iowa State University\\
dobson@iastate.edu}
}

\begin{document}
	
\maketitle	

\begin{abstract}
\looseness=-1
We develop LENORI, a Large Event Number of Outages Resilience Index measuring distribution system resilience with the number of forced line outages observed in large extreme events. LENORI is calculated from standard utility outage data. The statistical accuracy of LENORI is ensured by taking the logarithm of the outage data. A related Average Large Event Number of Outages metric ALENO is also developed, and both metrics are applied to a distribution system to quantify the power grid strength relative to the extreme events stressing the grid. The metrics can be used to track resilience and quantify the contributions of various types of hazards to the overall resilience.
\end{abstract}

\begin{IEEEkeywords}
Resilience metric, outages, blackouts, extreme events, weather, distribution system reliability, probability, risk
\end{IEEEkeywords}

\section{Introduction}

There are very many useful aspects of electric power system resilience, e.g. as described in \cite{StankovicPS23,DRWGreport,Pantelibook26}, but here we are concerned with outage resilience, which considers the power system strength with respect to extreme events in terms of the total number of resulting forced outages. In particular, we show how to quantify the outage resilience with new metrics that can be practically computed from the  outage  data currently collected by distribution utilities.
This is done with a novel {\bf L}arge {\bf E}vent {\bf N}umber of {\bf O}utages {\bf R}esilience {\bf I}ndex LENORI   measuring the resilience of the power system infrastructure in terms of the number of forced line outages observed in large extreme events.

\noindent
Distinctive attributes of LENORI are:
\begin{itemize}
\item LENORI mitigates the challenges of the rarity and variability of the number of outages in large resilience events by using a logarithmic transformation. This ensures that LENORI can be practically calculated with reasonable statistical accuracy from a few years of data.
\item LENORI responds to both the frequency and magnitude of extreme events.
\item LENORI can be readily calculated from the standard utility outage data that is used to calculate SAIDI.
\end{itemize}
We also introduce a related metric which is the {\bf A}verage {\bf L}arge {\bf E}vent {\bf N}umber of {\bf O}utages ALENO metric and show that LENORI is ALENO multiplied by the large event frequency.

\section{Literature review}
\looseness=-1
Quantitative resilience metrics often describe the dimensions, slopes, and areas of a performance curve, which tracks a resilience measure over time over the course of a resilience event \cite{Pantelibook26,NanRESS17,CarringtonPS21,IEEEwgs,EkishevaPMAPS22}. 
The metrics describe number of outages, number of customers disconnected, outage and restore rates, durations, nadirs, and the area under the resilience curve.
This paper only addresses the total number of outages in an event.

In distribution systems, the approximate power law in the probability distribution of number of outages in large events was noted in \cite{AhmadPS24}.
Power laws in other quantities correlated with the number of outages have also been observed, such as
customer minutes interrupted and customer costs \cite{AhmadPESL25,ahmadArxiv26a}.
Indeed, this paper has a useful parallelism with \cite{ahmadArxiv26a} in that \cite{ahmadArxiv26a} presents quite similar metrics based on customer minutes interrupted rather than number of outages.
In addition to describing a different facet of resilience, a salient difference with \cite{ahmadArxiv26a} is that the number of outages is a discrete quantity whereas customer minutes can be considered as continuous.

In transmission systems, the 2018 preprint \cite{DobsonArxiv18} discovered the power law in the distribution of the number of generations of outages, and proposed a System Event Propagation Index SEPSI to measure the absolute slope of the power law. 
Five years of comprehensive North American transmission line outage data collected by NERC show a power law in the distribution of the number of outages \cite{EkishevaPMAPS22}.

There are various types and severity of infrastructure damage, many of which cause a line outage. 
The forced line outages in an extreme event can be regarded as an indicator summarizing the amount of damage rather than directly listing the individual damaged components. 
Recording or estimating the various types of damage is of course valuable \cite{DRWGreport}, but it does not track the damage in a single quantity that is easily computed and analyzed,
whereas the number of line outages in an event can readily be extracted from routine utility records, and the subsequent calculation of metrics is straightforward.

\section{Utility data, outages,  and events}
Standard outage data recorded by distribution utilities include the start time and end time of each outage to the nearest minute and other information such as cause codes. The outage data is often collected by an outage management system. 
The outage data for this paper is a confidential dataset from a distribution utility in the Northeastern USA.

We extract from the outage data  resilience events in which outages bunch up and overlap in time according to the algorithm based on timing  in \cite{ahmadArxiv25}.
The events range in size from single outages to large events with tens or hundreds of outages caused by extreme weather. 
The number of outages in the  $i$th event is denoted by $N_i$.
There are some choices about which outages to count.
We only consider the forced outages, and momentary outages are included.

\section{Defining LENORI and ALENO} 
This section introduces the 
 {\bf L}arge {\bf E}vent {\bf N}umber of {\bf O}utages {\bf R}esilience {\bf I}ndex  LENORI and its associated {\bf A}verage {\bf L}arge {\bf E}vent {\bf N}umber of {\bf O}utages ALENO metric.

To define large events in terms of number of outages,
we set a threshold of the minimum number of outages in a large event as $N_{\rm L} $.
That is, a large event has a number of outages $N\ge N_{\rm L}$.
The way that $N_{\rm L} $ is chosen influences the interpretation and use of LENORI, and is discussed in section \ref{threshold}.

Suppose there are $n_{\rm large}$ large events over a period of $n_{\rm year}$ years,
with respective numbers of outages $N_1, N_2, ..., N_{n_{\rm large}}$.
Then we define  LENORI   as
    \begin{align}
    \mbox{LENORI}
   & = \frac{1}{n_{\rm year}} \sum_{i=1}^{n_{\rm large}}\ln \left(\frac{N_i}{N_{\rm L} -0.5 }\right)
   \label{LENORI}
\end{align}
The normalization by $n_{\rm year}$ in \eqref{LENORI} makes LENORI an annual index to allow comparison between metrics calculated from outage data collected over different numbers of years.

It is also useful to define the ALENO metric  as 
 \begin{align}
    \mbox{ALENO}
   & = \frac{1}{n_{\rm large}} \sum_{i=1}^{n_{\rm large}}\ln \left(\frac{N_i}{N_{\rm L} -0.5 }\right)
   \label{ALENO}
\end{align}
The annual frequency of large events is
\begin{align}
    f_{\rm large}= n_{\rm large}/n_{\rm year}
    \label{flarge}
    \end{align}
Then we can see from  (\ref{LENORI}),  (\ref{ALENO}), (\ref{flarge}) that 
\begin{align}
    \mbox{LENORI}
   & = f_{\rm large} \text{ ALENO}
   \label{LENORIALENO}
\end{align}
\looseness=-1
LENORI depends on both the frequency of large events and the size of large events.
In particular,  LENORI is directly proportional to large event frequency $f_{\rm large}$ whereas ALENO only responds to the size of large events. Indeed,
ALENO depends on the  average order of magnitude of large event size.

If the weather severity increases or the grid infrastructure weakens, the number of outages in individual large events increase and ALENO and LENORI increase. 
That is, these metrics respond to changes in the infrastructure strength relative to the weather severity.
In particular, if large events  have 
10\% more outages, then ALENO increases by $\ln  1.1=0.095$, and LENORI increases by 
$ f_{\rm large}(\ln 1.1)$.
If events just below the large event threshold $N_{\rm L}$ also have 10\% more outages,  then the increase in LENORI 
can be larger than $f_{\rm large}(\ln 1.1)$ because events with number of outages between 0.9$N_{\rm L}$ and $N_{\rm L}$ become large events.
Similarly, if a resilience investment reduces the number of outages in large events by 
10\%, then ALENO changes by  $\ln0.9=-0.105$ and LENORI changes by 
$ f_{\rm large}(\ln 0.9)$.

\section{LENORI and ALENO in more depth}

This section explains the basis for LENORI and ALENO in more depth by showing the heavy tails in the distribution of the number of outages, how ALENO characterizes the tail, and how logarithmic transformation mitigates the problem caused by the heavy tails of excessive variability in the metrics.
\subsection{Probability mass function of number of outages}

\begin{figure}
\centering
\includegraphics[width=0.9\linewidth]{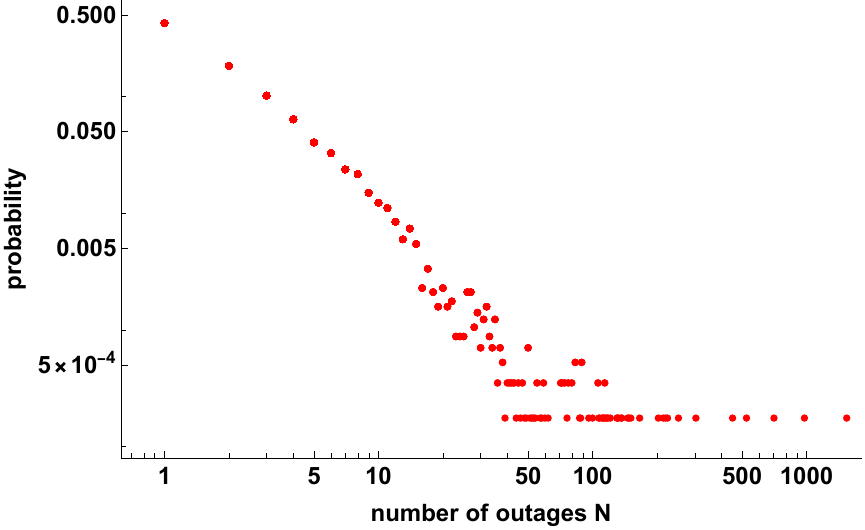}
    \caption{Log-log plot of probability mass function of number of outages.}
    \label{fig:pmf}
    \vspace{5mm}
\includegraphics[width=0.9\linewidth]{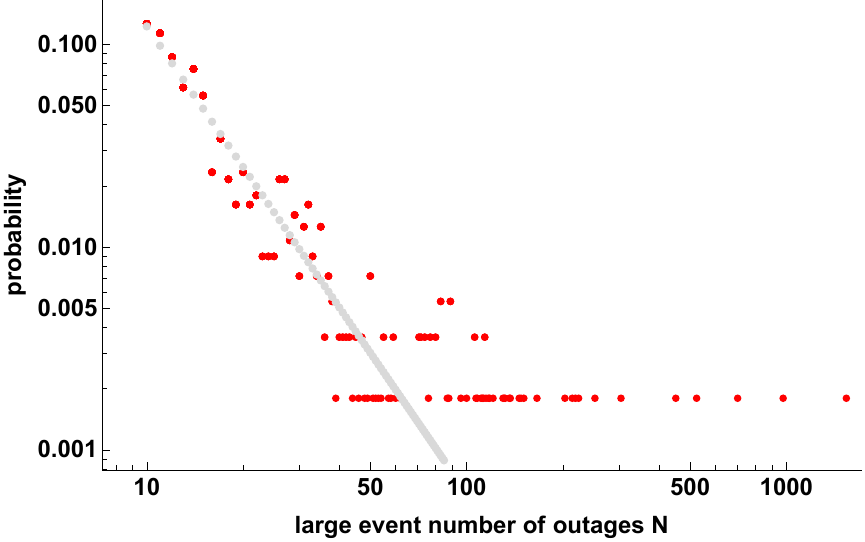}
    \caption{Tail of  probability mass function of Fig.~\ref{fig:pmf} (red dots) and its idealized power law \eqref{powerlaw} (light gray dots) with $\alpha$ estimated from the tail data. }
    \label{fig:pmftail}
\end{figure}

Fig.~\ref{fig:pmf} shows the log-log plot of the probability mass function of the number of outages in all events for the distribution system.
To determine the resilience and the statistical behavior, we now  focus on the large events in the tail of this distribution.

\subsection{Linear approximation to the power law tail}
\label{linear}

This subsection relates ALENO to the slope of the tail of the probability mass function for the number of outages.

Since we are interested in characterizing the tail of the probability mass function, we select only the tail event data with $N\ge N_{\rm L}$.
The empirical distribution of $N$ conditioned on $N\ge N_{\rm L}$ is plotted on a log-log scale in Fig.~\ref{fig:pmftail}.
The tail in Fig.~\ref{fig:pmftail} is approximately a straight line, and thus has approximate power-law behavior.

The exact idealized power law behavior is described  by a discrete power law distribution $P$ with  probability mass function 
\begin{flalign}
    f_P(n)=\frac{1}{\zeta(\alpha+1,N_{\rm L})}
   n^{-(\alpha+1)},\quad
      n=N_{\rm L},N_{\rm L}\!+\!1,...
     \label{powerlaw}&
     \end{flalign}
where  $
    \zeta(\alpha+1,N_{\rm L})=\sum_{n=N_{\rm L}}^\infty n^{-(\alpha+1)}$
is the Hurwitz zeta function. The distribution \eqref{powerlaw} is sometimes called a zeta or Zipf or discrete Pareto distribution.
The slope magnitude of \eqref{powerlaw} on the log-log plot in Fig.~\ref{fig:pmftail} is $\alpha+1$, as can be verified by taking the logarithm of \eqref{powerlaw}. 
$\alpha$ is called the tail index\footnote{For a continuous distribution the tail index $\alpha$ is the absolute slope of the probability exceedance function (CCDF); for the discrete distribution \eqref{powerlaw}, $\alpha$ is the asymptotic absolute slope of its probability exceedance function.}.

Then, according to Clauset \cite{ClausetSIAM09},
the maximum likelihood estimate of the tail index $\alpha$ is 
well approximated by 
 \begin{align}
 \alpha
   & = \left[\frac{1}{k} \sum_{i=1}^{k}\ln \left(\frac{N_i}{N_{\rm L} -0.5 }\right)\right]^{-1}
   =\frac{1}{\mbox{ALENO}}
   \label{alphaclauset}
\end{align}
for $N_{\rm L} \ge 6$.
This important relation between ALENO and the tail index $\alpha$ shows how ALENO describes the tail of the probability mass function
of direct interest to resilience risk. 
The tail index and ALENO capture the linear trend of the large event sizes, and this trend, if extrapolated, governs the  blackouts that are larger than the largest blackout that has so far been observed.
There is considerable inherent uncertainty in extrapolating beyond the largest observed blackout, but given the dominant risk of the largest possible blackouts, it does seem better to try to approximately account for this risk rather than ignoring it. 
Since ${\rm ALENO}$ is a factor of ${\rm LENORI}$ as shown in \eqref{LENORIALENO}, ${\rm LENORI}$ also incorporates the linear trend of the large event sizes on the log-log plot.

\subsection{Log transformation converts heavy tail to light tail}

\begin{figure*}[ht]
    \centering
   \includegraphics[width=0.9\linewidth]{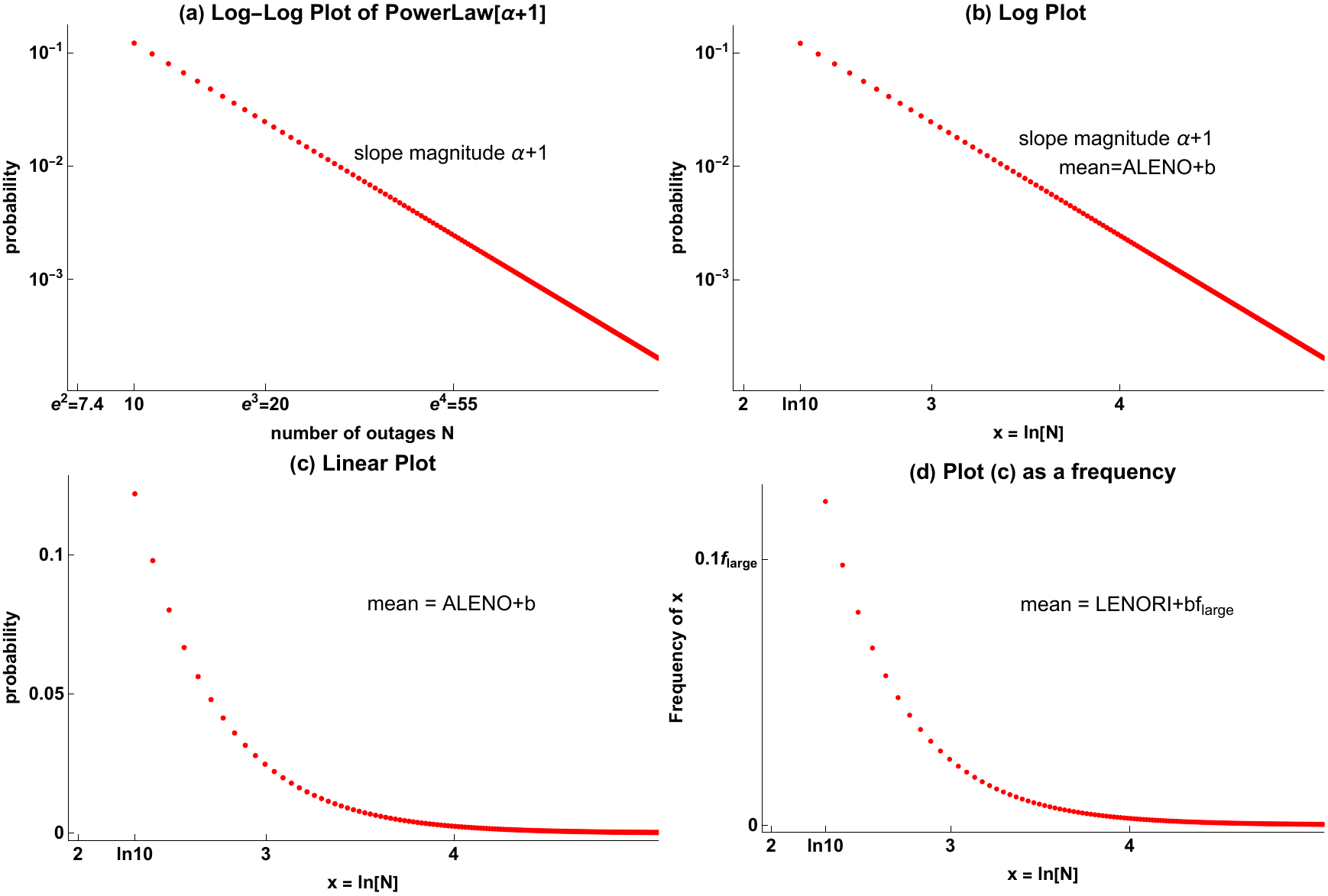}
    \caption{Plot (a) shows on a log-log plot the idealized Pareto probability mass function \eqref{powerlaw} of the event number of outages $N$ for the large event tail $N\ge N_{\rm L}=10$. The tail index $\alpha$ characterizes the tail since the slope magnitude of the tail is $\alpha+1$.
    Plot (b) applies a logarithm to the same number of outages data as plot (a) by relabeling the horizontal axis. Since plot (b) is a probability mass function that is the same straight line of slope magnitude $\alpha+1$ but now on a log plot, the log-transformed data $X=\ln N$ is a light-tailed distribution similar to an exponential distribution with parameter $\alpha+1$ and mean ALENO$+b$, where $b=\ln(N_{\rm L}-0.5)$. Plot (c) is the same as plot (b) but now on a linear plot.
    Plot (d) is the same as plot (c) except that the vertical axis is rescaled to show frequency. The annual large event frequency is $f_{\rm large}$. The mean of the frequency function is LENORI$+bf_{\rm large}$.
    }
    \label{fig:logplots}
\end{figure*}

If we take logarithm of the idealized tail data and define $X=\ln P$ then $X$ has the following probability mass function with parameter $\alpha+1$:
\begin{align}
f_X(x)
    &\!=\!\frac{1}{\zeta(\alpha\!+\!1,N_{\rm L})}
   e^{-(\alpha\!+\!1)x},\,x\!=\!\ln\! N_{\rm L},\ln(N_{\rm L}\!+\!1),...
    \label{geom}
\end{align} 
$X$ can be regarded as a discrete form of the exponential distribution with support on the logarithms of integers $\ge N_L$.
Since exponential distributions are light tailed, 
this shows that the logarithm of the heavy-tailed power law data with slope magnitude $\alpha+1$ has a light-tailed distribution with parameter $\alpha+1$; also see Fig. \ref{fig:logplots}. 

The first two moments of $X$ are, for $k=1,2$,
\begin{align}
    {\rm E}[X^k]={\rm E}[(\ln P)^k]
    =\frac{1}{\zeta(\alpha+1,N_{\rm L})}\sum_{n=N_{\rm L}}^\infty &
   (\ln n)^k n^{-(\alpha+1)}\notag
\end{align}

\subsection{Required number of large events and metric variability}

 We now estimate the statistical variation of LENORI with its Relative Standard Error ${\rm RSE}_{\rm LEN}$,
which is the standard deviation of LENORI divided by the mean of LENORI. 
The statistical variation of LENORI arises from the variability  in the number of large events and the variability in the magnitude of large events.
Let the number of large events be the random variable 
$N_{\rm large}$.
The statistical variation of $N_{\rm large}$ is modeled as  Poisson[$n_{\rm large}$]  independent of the magnitude of large events. It follows that ${\rm E}N_{\rm large}={\rm Var}N_{\rm large}=n_{\rm large}$.

For the statistical variation in the magnitude of large events, we approximate the tail of the distribution with a straight line on the log-log plot so that it follows the Pareto distribution (\ref{powerlaw}).
Then LENORI and ALENO can be considered as random variables that combine independent and identically distributed samples $X_1$, $X_2$, ..., $X_{N_{\rm large}}$ from the 
log-transformed data (\ref{geom}). 
Let $b=\ln (N_{\rm L}-0.5)$ and let
\begin{align}
  L=n_{\rm year}\,{\rm LENORI}= \sum_{i=1}^{N_{\rm large}}(X_i-b)
\end{align}
The Wald equation gives ${\rm E}L=n_{\rm large}({\rm E}X-b)$ and
the Blackwell-Girshick equation gives 
\begin{align}
    {\rm Var}L&={\rm E}N_{\rm large}{\rm Var}(X-b)+({\rm E}[X-b])^2{\rm Var}N_{\rm large}\\
    &=n_{\rm large}{\rm E}[(X-b)^2]
\end{align}
\begin{flalign}
\text{\rm Then} \hspace{1 cm}  {\rm RSE}_{\rm LEN} &= {\rm RSE}_{L} =\frac{\sqrt{{\rm E}[(X-b)^2]}}{({\rm E}X-b)\sqrt{n_{\rm large}}}&
  \label{RSELENORI}
\end{flalign}
\begin{flalign}
\text{\rm Also} \hspace{1 cm}
      {\rm ALENO}&=\frac{1}{N_{\rm large}}
    \sum_{i=1}^{N_{\rm large}}(X_i -b)&\\
{\rm RSE}_{\rm ALE}
&=\frac{\sigma(X)}{({\rm E}X-b)\sqrt{ n_{\rm large}}}&
\end{flalign}

The statistical variability of LENORI calculated from $n_{\rm year}$ years of data is governed by the number $n_{\rm large}$ of large events in the $n_{\rm year}$ years.
According to \eqref{RSELENORI},
to ensure for LENORI a statistical accuracy ${\rm RSE}_{\rm LEN}^{\rm max}$ so that 
 ${\rm RSE}_{\rm LEN}\le{\rm RSE}_{\rm LEN}^{\rm max}$,
 the minimum number of large events required is 
 \begin{align}
   n_{\rm large}^{\rm min}=  
    \frac{{\rm E}[(X-b)^2]}{({\rm E}X-b)^2({\rm RSE}_{\rm LEN}^{\rm max})^2}
    \label{nlargemin}
\end{align}
 Then the minimum number of years needed for $n_{\rm large}^{\rm min}$ large events is 
 \begin{align}
   n_{\rm year}^{\rm min}= 
    n_{\rm large}^{\rm min}/f_{\rm large}^{\rm all}
    \label{RSEcalc2}
\end{align}
where  $f_{\rm large}^{\rm all}$ is the annual frequency of large events using all the observed years of data.

In this paper, we adopt ${\rm RSE}_{\rm LEN}^{\rm max}\approx 0.1$, implying that the standard deviation of an estimate of LENORI is no more than 10\% of its mean value.
Under reasonable normality assumptions, this indicates that $\sim$68\% of the estimates of LENORI lie within 10\% of the mean, and that $\sim$90\% of the estimates of LENORI lie within 16\% of the mean.

The value of $n_{\rm year}^{\rm min}$ implies that LENORI can be tracked with reasonable statistical accuracy over time with an $n_{\rm year}^{\rm min}$ year sliding window. Table~\ref{utilitydata} shows that this is practical for our data with $n_{\rm year}^{\rm min}\approx 2$.

\subsection{Variability of LENORI with no logarithm}

We first to try to evaluate ${\rm RSE}_P$.
Using \eqref{powerlaw}, the moments of $P$ are 
\begin{align}
    {\rm E}[P^k]
    =\frac{1}{\zeta(\alpha+1,N_{\rm L})}\sum_{n=N_{\rm L}}^\infty 
    n^k n^{-(\alpha+1)}
    \label{EPk}
\end{align}
For $k=1$, the mean ${\rm E} P$ is infinite for $\alpha\le 1$.
The mean ${\rm E} P$ exists for $\alpha>1$ but is difficult to practically compute from limited data for $1<\alpha<1.5$.
However, for $k=2$ and $\alpha\le 2$ as in our data, \eqref{EPk} implies that the variance and standard deviation of $P$ are infinite, and hence ${\rm RSE}_P$ is infinite.

\looseness=-1
However, this analysis using $P$ is not realistic because \eqref{powerlaw} indicates  an unbounded number of outages $N$, whereas in practice $N$ is bounded by a maximum number of outages $N_{\rm max}$ so that  $N\le N_{\rm max}$. $N_{\rm max}$ corresponds to a blackout of the entire system.
For our case, the maximum observed blackout has $N_{\rm maxobs}=1540$ outages, and we roughly estimate the largest possible blackout as $N_{\rm max}=5000$.

The power law distribution \eqref{powerlaw} conditioned on $N \le N_{\rm max}$ is the bounded distribution $Pb$ with probability mass function $f_{Pb}(n)=f_{P}(n)/c$, $ n=N_{\rm L},N_{\rm L}\!+\!1,...,N_{\rm max}$, where the renormalization constant\footnote{$c=0.9997$ is close enough to one that it can be neglected.} $c=1-\zeta(\alpha+1,N_{\rm max}+1)/\zeta(\alpha+1,N_{\rm L})$.
The first two moments of $Pb$ for use in computing ${\rm RSE}_{Pb}=\sqrt{{\rm E}[Pb^2]-({\rm E}[Pb])^2}/{\rm E}[Pb]$ are 
\begin{align}
{\rm E}[Pb^k]=\frac{1}{c\,\zeta(\alpha+1,N_{\rm L})}\sum_{n=N_{\rm L}}^{N_{\rm max}}
    n^k n^{-(\alpha+1)},\  k=1,2.
\end{align}

Now  remove the logarithm from \eqref{LENORI} to consider an index 
\begin{align}
{\rm LENnolog}=\frac{1}{n_{\rm year}} \sum_{i=1}^{n_{\rm large}} \frac{N_i}{N_{\rm L} -0.5 }
\end{align}
Since RSE does not depend on scaling and using Appendix B of \cite{ahmadArxiv26a},  
\begin{align}
{\rm RSE}_{\rm LENnolog}&=
{\rm RSE}\left[\sum_{i=1}^{N_{\rm large}} N_i\right]=\frac{\sqrt{1+({\rm RSE}_{Pb})^2}}{\sqrt{n_{\rm large}}}
     \label{RSEALENOnolog}
\end{align}
 Let $n_{\rm large}^{\rm minnolog}$ be the minimum number of large events needed for ${\rm LEN}_{\rm nolog}$ to have statistical accuracy  
${\rm RSE}_{\rm LEN nolog}^{\rm max}=0.1$ the same as LENORI.  Then we can calculate
$  n_{\rm large}^{\rm minnolog}=
    (1+({\rm RSE}_{Pb})^2)/0.01$.
    
Table~\ref{utilitydata} shows $n_{\rm large}^{\rm minnolog}$ and $n_{\rm large}^{\rm min}$. 
For the same statistical accuracy, LENnolog requires five times more large events than LENORI and more than 11 years to gather enough large events.
Thus LENnolog requires an excessively large amount of data for tracking resilience with this utility data.

\subsection{Large event threshold $N_{\rm L}$ }
\label{threshold}

For setting the large event threshold $N_L$ in our case, there is a tradeoff between the statistical accuracy of LENORI and the fit of the linear approximation to the large event power law tail in subsection~\ref{linear}.
A good enough linear fit enables the useful interpretation of ${\rm ALENO}^{-1}$ as estimating the tail index of the PMF of the number of outages.
This interpretation is significant not only for describing the risk of the largest observed events, but also, when this linear trend is extrapolated, giving the best available description (even if subject to uncertainties) of 
the very substantial risk of blackouts that are larger than those already observed.

\begin{figure}[h]
    \centering
   \includegraphics[width=0.9\linewidth]{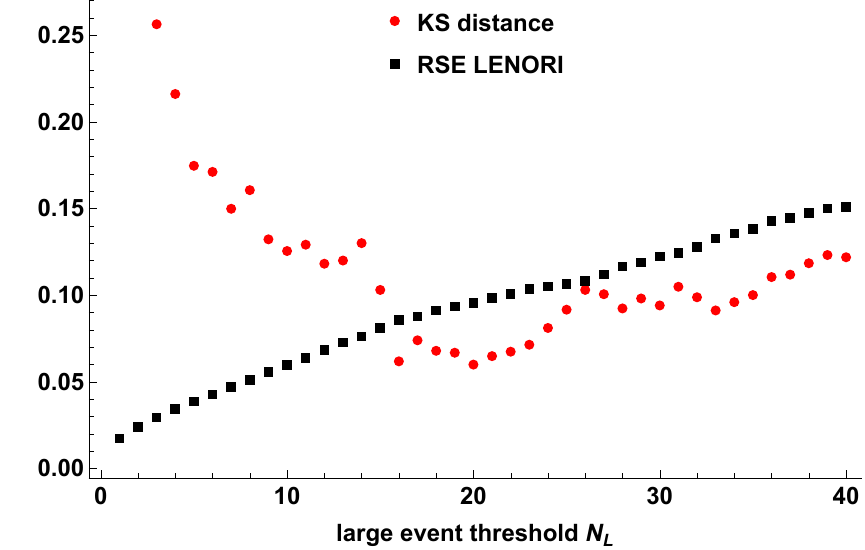}
    \caption{Kolmogorov–Smirnov distance between power law fit and event number of outages and LENORI Relative Standard Error as large event threshold $N_L$ varies. There is a tradeoff in minimizing both functions.
    }
    \label{fig:KSandRSEHEN}
\end{figure}

The statistical accuracy of LENORI can be measured by its Relative Standard Error ${\rm RSE_{\rm LEN}}$ in (\ref{RSELENORI}). As $N_L$ increases, the number of large events $n_{\rm large}$ decreases and ${\rm RSE_{\rm LEN}}$ increases as shown in Fig.~\ref{fig:KSandRSEHEN}. 
The fit of the linear approximation to the large event power law tail can be measured by the Kolmogorov–Smirnov distance between the estimated Pareto distribution \eqref{powerlaw} and the empirical distribution of the number of outages in the large events. 
If we only optimized the fit of the linear approximation, then this would amount to applying Clauset’s method \cite{ClausetSIAM09}, and the best fit is at the minimum of the Kolmogorov–Smirnov distance at $N_L=20$, or very close to the minimum at $N_L=16$ as shown in Fig.~\ref{fig:KSandRSEHEN}.
So the best fit requires increasing $N_L$ to at least 16,  but limiting ${\rm RSE_{\rm LEN}}$ favors a smaller $N_L$. To achieve ${\rm RSE_{\rm LEN}}<0.1$ with some margin to allow for decomposing LENORI by cause or season, here we choose $N_L=10$.
While LENORI is shown to work well in this paper, 
the main limitation to its adoption in practice is 
the lack of experience in choosing $N_L$ in other power system cases. 
Future work will examine other cases to find a general approach to choosing $N_L$.

\section{Results}

Table~\ref{utilitydata} shows the utility data, the metrics ALENO and LENORI, and their statistical accuracies, calculated for all six years of data.
\vspace{-2mm}

\begin{table}[hbpt]
	\caption{Utility data and metrics} 
	\label{utilitydata}
	\centering
    \vspace{-2mm}
\begin{tabular}{rl@{\hspace{1cm}}rl} 
 \text{$\alpha$} & 1.3 &\text{$n_{\rm year}$} & 6 \\
 \text{ALENO} & 0.769 &\text{$n_{\rm large}^{\rm min}$} & 199.\\
 \text{LENORI} & 71.5 &\text{$n_{\rm year}^{\rm min}$} & 2.14\\
 \text{${\rm RSE}_{\rm ALE}$} & 0.0421 &\text{$n_{\rm large}^{\rm minnolog}$} & 1090.  \\
 \text{${\rm RSE}_{\rm LEN}$} & 0.0597 & \text{$n_{\rm year}^{\rm minnolog}$} & 11.7\\
 \text{${\rm RSE}_{Pb}$} & 3.15 &\text{$N_{\rm maxobs}$} & 1540 \\
 \text{$n_{\rm large}$} & 558 & \text{$N_{\rm max}$} & 5000\\
 \text{$f_{\rm large}$} & 93. &\text{$N_{\rm L}$} & 10 
\end{tabular}
\end{table}

\begin{table}[hbpt]
	\caption{Decomposing resilience  by season} 
	\label{season}
	\centering
\begin{tabular}{rccc} &all&summer&non-summer\\\hline
 \text{$\alpha$} & 1.3 & 1.36 & 1.23 \\
 \text{ALENO} & 0.769 & 0.737 & 0.816 \\
 \text{LENORI} & 71.5 & 41.3 & 30.2 \\
 \text{${\rm RSE}_{\rm ALE}$} & 0.0421 & 0.0542 & 0.0666 \\
 \text{${\rm RSE}_{\rm LEN}$} & 0.0597 & 0.0769 & 0.0945 \\
 \text{$n_{\rm large}$} & 558 & 336 & 222 \\
 \text{$n_{\rm large}^{\rm min}$} & 199. & 199. & 198. \\
 \text{$f_{\rm large}$} & 93. & 56. & 37. \\
 \text{$n_{\rm year}$} & 6 & 6 & 6 
\end{tabular}
\end{table}

\begin{table}[hbpt]
	\caption{Decomposing resilience  by cause} 
	\label{cause}
	\centering
\begin{tabular}{rcccc} &all&tree&weather&other\\\hline
  \text{$\alpha$} & 1.3 & 1.03 & 0.503 & 2.51 \\
 \text{ALENO} & 0.769 & 0.972 & 1.99 & 0.399 \\
 \text{LENORI} & 71.5 & 45.3 & 9.61 & 16.5 \\
 \text{${\rm RSE}_{\rm ALE}$} & 0.0421 & 0.0588 & 0.168 & 0.0628 \\
 \text{${\rm RSE}_{\rm LEN}$} & 0.0597 & 0.0839 & 0.25 & 0.0892 \\
 \text{$n_{\rm large}$} & 558 & 280 & 29 & 249 \\
 \text{$n_{\rm large}^{\rm min}$} & 199. & 197. & 182. & 198. \\
 \text{$f_{\rm large}$} & 93. & 46.7 & 4.83 & 41.5 \\
 \text{$n_{\rm year}$} & 6 & 6 & 6 & 6 
\end{tabular}
\end{table}

The resilience due to different causes can be assessed by applying the metrics to subsets of the data.
Table~\ref{season} shows how the distribution system resilience depends on whether the season is summer, where summer is defined to be the months of June, July, August, September.
ALENO indicates that non-summer events are a bit larger on average than summer events. But the large events are more frequent in summer (see $f_{\rm large}$), so that overall,
LENORI indicates that summer months are less resilient than non-summer months. Since summer and non-summer account for all the large events, the summer and non-summer LENORI add to LENORI for all year.

The distribution system outage cause codes can be grouped into cause codes related to trees, or weather, or other causes. Then the cause code for an event is the majority cause code.
Of course trees are often involved in bad weather, as, for example,  when high winds detach parts of trees and they fly into lines, but 
we are interested to consider them separately to assess the importance of tree trimming for resilience.
Table~\ref{cause} shows how the distribution system resilience depends on tree, weather, or other cause codes.
ALENO shows that the individual large events caused by weather are significantly more severe than those caused by trees, but since $f_{\rm large}$ shows that the tree large events are even more frequent, the overall effect shown by LENORI is that tree large events affect resilience more. 
Thus tree trimming that reduces large tree event outages by 10\% will be more effective in reducing the number of outages than hardening that reduces large weather event outages by 10\%. There is some loss of statistical accuracy in the ALENO and LENORI metrics for the limited number 29 of weather events, but this would be much worse if the logarithmic transformation was not used.

Table~\ref{tracking} shows that tracking the yearly resilience of the distribution system over a window of 2 years is practical. In particular, the statistical accuracy of LENORI is maintained at ${\rm RSE}_{\rm LEN}\approx 0.1$ for biennial tracking.

\begin{table}[hbpt]
	\caption{Biennial tracking of resilience with LENORI and ALENO} 
	\label{tracking}
	\centering
\begin{tabular}{rcccc} 
&2012&2013&2014&2015\\
&2013&2014&2015&2016\\\hline
\text{$\alpha$} & 1.31 & 1.38 & 1.47 & 1.59 \\
 \text{ALENO} & 0.766 & 0.726 & 0.682 & 0.629 \\
 \text{LENORI} & 75.1 & 72.6 & 71.3 & 71.7 \\
 \text{${\rm RSE}_{\rm ALE}$} & 0.071 & 0.0703 & 0.0688 & 0.0659 \\
 \text{${\rm RSE}_{\rm LEN}$} & 0.101 & 0.0997 & 0.0976 & 0.0934 \\
 \text{$n_{\rm large}$} & 196 & 200 & 209 & 228 
\end{tabular}
\end{table}

\section{Discussion and Conclusion}

We examine the probability distribution of the number of forced outages in large resilience events in a distribution system.
There is heavy-tailed power law behavior. We propose LENORI and ALENO resilience metrics based on the number of  outages in large events that describe the power grid strength relative to the weather severity and frequency. 
These metrics mitigate the problem of excessive statistical variability due to the heavy tails by taking the logarithm of the number of outages. 
While the number of outages in each event is a useful metric for individual events, it varies too erratically 
to characterize the power system resilience. 
Indeed, LENORI without the logarithm requires too much data ($>11$ years) to have reasonable statistical accuracy.
As regards metrics based on customer minutes interrupted, SAIDI is well known to be erratic when major event days are included, and this analogous problem is solved by the logarithmic metric SALEDI \cite{ahmadArxiv26a}.

\looseness=-1
ALENO also determines the slope magnitude tail index $\alpha$  that characterizes the large event trend.
The heavy tail for the distribution of the number of outages is less heavy ($\alpha=1.3$)  than the tail for the distribution of customer minutes interrupted ($\alpha=0.83$) reported in \cite{ahmadArxiv26a} for the same utility.
The extent of high statistical variability of metrics calculated  from the number of outages without a logarithm depends on the tail index $\alpha$, so we consider whether the value of $\alpha=1.3$ in our data is typical.
Initial calculations for four other USA distribution systems give $\alpha$ ranging from 1.0 to 1.6 with mean value 1.2, giving some indication that our data is typical and that the logarithm is needed to limit metric variability in other cases.
Future work should confirm this by systematically testing the new metrics on a wide range of systems and optimizing the choice of the large event threshold $N_{\rm L}$.

\looseness=-1
The number of forced line outages in a power system resilience event is a basic metric describing the event size that summarizes the infrastructure damage in the event.
For resilience, we focus on the large events, and describe with metrics  the tail of the probability mass function of 
the numbers of outages.
However, for our distribution system data that has heavy tails, simply adding or averaging the number of outages in large events 
to obtain a metric does not work well because too many large events need to be observed over many years to obtain reasonable statistical accuracy. 
This problem is mitigated by adding or averaging the logarithm of the number of outages in large events to obtain the LENORI and ALENO metrics. 
The ALENO metric determines the tail index $\alpha$, and the LENORI metric also accounts for the frequency of large events. 
The LENORI and ALENO metrics have sufficiently low statistical variability that they can be used to track the power grid resilience over the years, and to be decomposed by season or cause to quantify the contributions of the various types of hazards to the overall resilience. 
This can be used to prioritize  grid-hardening resilience investments for these types of hazards in terms of their resilience benefits.

\printbibliography

\end{document}